\documentclass[twocolumn,english]{article}
\usepackage[T1]{fontenc}
\usepackage[latin9]{inputenc}
\usepackage{babel}
\usepackage{float}
\usepackage{textcomp}
\usepackage{amstext}
\usepackage{graphicx}
\usepackage[unicode=true,pdfusetitle,
 bookmarks=true,bookmarksnumbered=false,bookmarksopen=false,
 breaklinks=false,pdfborder={0 0 1},backref=false,colorlinks=false]
 {hyperref}

\makeatletter

\providecommand{\tabularnewline}{\\}
\floatstyle{ruled}
\newfloat{algorithm}{tbp}{loa}
\providecommand{\algorithmname}{Algorithm}
\floatname{algorithm}{\protect\algorithmname}

     \usepackage{usenix} 

\@ifundefined{date}{}{\date{}}
\makeatother

\usepackage{listings}

\begin{document}

\title{The Spy in the Sandbox -- Practical Cache Attacks in Javascript}

\author{Yossef Oren, Vasileios P. Kemerlis, Simha Sethumadhavan and Angelos
D. Keromytis\\
Computer Science Department, Columbia University\\
\{yos | vpk | simha | angelos\}@cs.columbia.edu}
\maketitle
\begin{abstract}
We present the first micro-architectural side-channel attack which
runs entirely in the browser. In contrast to other works in this genre,
this attack does not require the attacker to install any software
on the victim's machine -- to facilitate the attack, the victim needs
only to browse to an untrusted webpage with attacker-controlled content.
This makes the attack model highly scalable and extremely relevant
and practical to today's web, especially since most desktop browsers
currently accessing the Internet are vulnerable to this attack. Our
attack, which is an extension of the last-level cache attacks of Yarom
et al. \cite{Yarom_LGHL_15}, allows a remote adversary recover information
belonging to other processes, other users and even other virtual machines
running on the same physical host as the victim web browser. We describe
the fundamentals behind our attack, evaluate its performance using
a high bandwidth covert channel and finally use it to construct a
system-wide mouse/network activity logger. Defending against this
attack is possible, but the required countermeasures can exact an
impractical cost on other benign uses of the web browser and of the
computer.
\end{abstract}

\section{Introduction}

Side channel analysis is a remarkably powerful class of cryptanalytic
attack. It lets attackers extract secret information hidden inside
a secure device by analyzing the physical signals (power, radiation,
heat, etc.) the device emits as it performs a secure computation \cite{DBLP:books/daglib/0017272}.
Allegedly used by the intelligence community as early as World War
II, and first discussed in an academic context by Kocher et al. in
1996 \cite{DBLP:conf/crypto/Kocher96}, side channel analysis has
been shown to be effective in breaking into myriad real-world systems,
from car immobilizers to high-security cryptographic coprocessors
\cite{DBLP:conf/crypto/EisenbarthKMPSS08,DBLP:conf/ches/OswaldP11}.
A particular kind of side-channel attack which is relevant to personal
computers is the cache attack, which exploits the use of cache memory
as a shared resource between different processes or users to disclose
secret information \cite{DBLP:conf/ctrsa/OsvikST06,DBLP:conf/sp/Hu92}. 

While the potency of side-channel attacks is established without question,
their application to practical systems is relatively limited. The
main limiting factor to the practicality of side-channel attacks is
the problematic \textbf{attack model} they assume: with the exception
of network-based timing attacks, most side-channel attacks require
that the attacker be in close proximity to the victim. Cache attacks,
in particular, typically assume that the attacker is capable of executing
arbitrary binary code on the victim's machine. While this assumption
holds for Infrastructure/Platform-as-a-Service (IaaS/PaaS) environments
such as Amazon's cloud computing platform, it is less relevant for
other settings.

In this report we challenge this limiting security assumption by presenting
a successful cache attack which assumes a far more relaxed and practical
attacker model. In our attacker model, the victim merely has to \textbf{access
a website} owned by the attacker. Despite this minimal attack model,
we show how the attacker can still launch an attack in a practical
time frame and extract meaningful information from the system under
attack. Keeping in tune with this computing setting, we chose to focus
our attacks not on cryptographic key recovery but rather on \textbf{tracking
user behavior}. The attacks described in this report are therefore
highly practical: practical in the assumptions and limitations they
cast upon the attacker; practical in the time they take to run; and
practical in terms of the benefit they deliver to the attacker. To
the best of our knowledge, this is the first side-channel attack which
can scale effortlessly into millions of targets. 

For our attacks we assume that the victim is using a personal computer
powered by a late-model Intel CPU. We furthermore assume that the
user is accessing the web through a browser with comprehensive HTML5
support. As we show in Subsection \ref{sub:Is-Your-Computer}, this
covers a vast majority of personal computers connected to the Internet.
The victim is coerced to view a webpage containing an attacker-controlled
element such as an advertisement. The attack code itself, which we
describe in more detail in Section \ref{sec:Attack-Methodology},
executes a Javascript-based \textbf{cache attack}, which allows it
to track accesses to the DUT's last-level cache (LLC) over time. Since
this single cache is shared by all CPU cores and by all users, processes
and protection rings, this information can provide the attacker with
a detailed knowledge of the user and the system under attack.

\subsection{\label{sub:The-Memory-Architecture}The Memory Architecture of Modern
Intel CPUs}

\begin{figure*}
\begin{centering}
\includegraphics[width=0.6\textwidth]{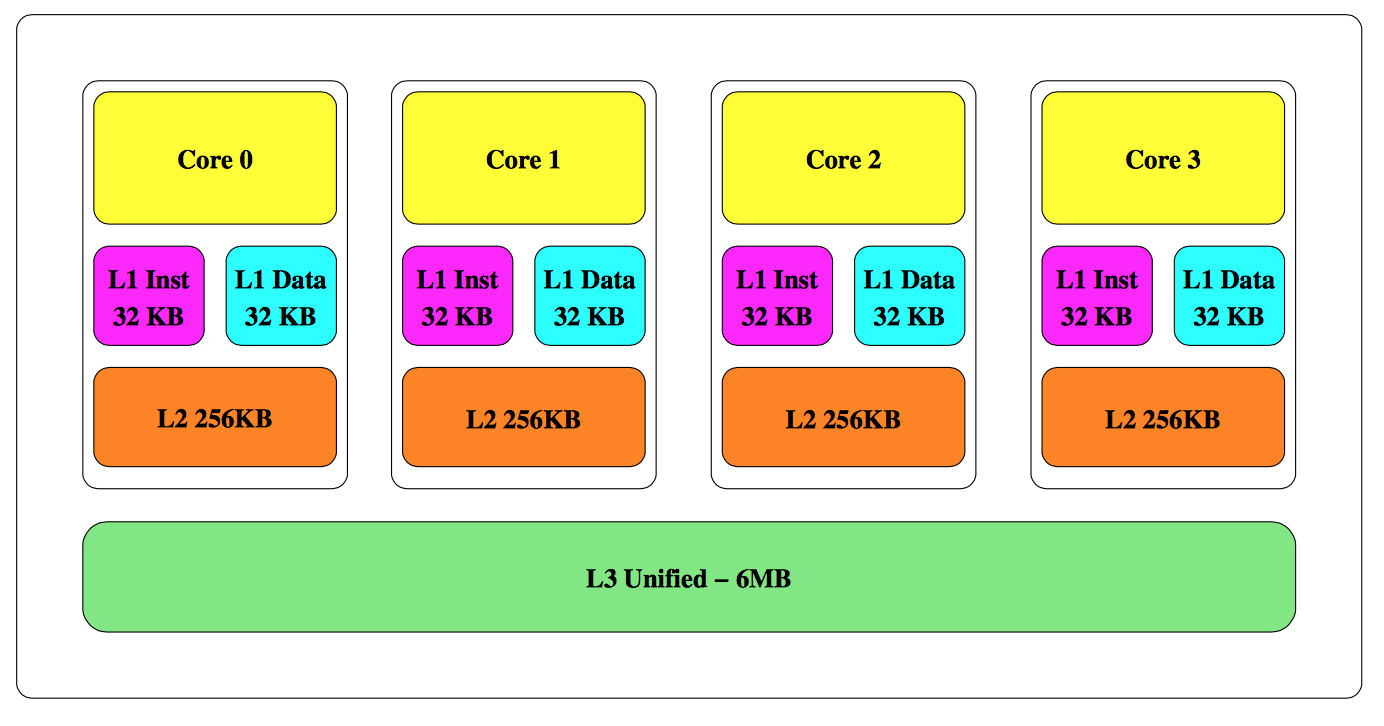}
\par\end{centering}

\protect\caption{\label{fig:The-Intel-Ivy}The Intel Ivy Bridge Cache Architecture
(taken from \cite{DBLP:conf/uss/YaromF14})}
\end{figure*}

Modern computer systems typically incorporate a high-speed central
processing unit (CPU) and a large amount of lower-speed random access
memory (RAM). To bridge the performance gap between these two components,
modern computer systems make use of \textbf{cache memory} -- a type
of memory element with a smaller size but a higher performance, which
contains a subset of the RAM which has been recently accessed by the
CPU. The cache memory is typically arranged in a \textbf{cache hierarchy},
with a series of progressively larger and slower memory elements being
placed in \textbf{levels} between the CPU and the RAM. Figure \ref{fig:The-Intel-Ivy},
taken from \cite{DBLP:conf/uss/YaromF14}, shows the cache hierarchy
used by Intel Ivy Bridge series CPUs, incorporating a small, fast
\textbf{level 1 (L1) cache}, a slightly larger \textbf{level 2 (L2)
cache}, and finally a larger \textbf{level 3} \textbf{(L3) cache }which
is then connected to the RAM. The current generation of Intel CPUs,
code named Haswell, extends this hierarchy by another level of embedded
DRAM (eDRAM), which is not discussed here. Whenever the CPU wishes
to access a memory element, the memory element is first searched for
in the cache hierarchy, saving the lengthy round-trip to the RAM.
If the CPU requires an element which is not currently in the cache,
an event known as a \textbf{cache miss, }one of the elements currently
residing in the cache must be \textbf{evicted }to make room for this
new element.

The Intel cache micro-architecture is \textbf{inclusive} -- all elements
in the L1 cache must also exist in the L2 and L3 caches. Conversely,
if a memory element is evicted from the L3 cache, it is also immediately
evicted from the L2 and L1 cache. It should be noted that the AMD
cache micro-architecture is exclusive, and thus the attacks described
in this report are not immediately applicable to that platform.

This report focusses on the level 3 cache, commonly referred to as
the last-level cache (LLC). Due to the LLC's relatively large size,
it is not efficient to search its entire contents whenever the CPU
accesses the memory. Instead, the LLC is divided into \textbf{cache
sets}, each covering a fixed subset of the memory space. Each of these
cache sets contains several \textbf{cache lines}. For example, the
Intel Core i7-3720QM processor, which belongs to the Haswell family,
includes $8192=2^{13}$ cache sets, each of which can hold 12 lines
of $64=2^{6}$ bytes each, giving a total cache size of 8192x12x64=6MB.
When the CPU  needs to check whether a given physical address is present
in the L3 cache, it calculates which cache set is responsible for
this address, then only checks the cache lines corresponding to this
set. As a consequence, a cache miss event for a physical address can
result in the eviction of only one of the relatively small amount
of lines sharing its cache set, a fact of which we make great use
in our attack. The method of mapping between 64-bit physical addresses
and 13-bit cache set indices has been reverse engineered by Hund et
al. in 2013 \cite{DBLP:conf/sp/HundWH13}: of the 64 physical address
bits, bits 5 to 0 are ignored, bits 16 to 6 are taken directly as
the lower 11 bits of the set index, and bits 63 to 17 are hashed to
form the upper 2 bits of the cache index. The LLC is shared between
all cores, threads, processes, users, and even virtual machines running
on a certain CPU chip, regardless of privilege rings or other protection
similar mechanisms.

Modern personal computers use a \textbf{virtual memory mechanism},
in which user processes do not typically have direct knowledge or
access to the system's physical memory. Instead, these processes are
allocated virtual memory \textbf{pages}. When a virtual memory page
is accessed by a currently executing process, the operating system
dynamically associates the page with a \textbf{page frame} in physical
memory. The CPU's memory management unit (MMU) is in charge of mapping
between the virtual memory accesses made by different processes and
accesses to physical memory. The size of pages and page frames in
most Intel processors is typically set to 4KB, and both pages and
page frames are \textbf{page aligned} -- the starting address of each
page is a multiple of the page size. This means that the lower 12
bits of any virtual address and its corresponding virtual address
are generally identical, another fact we use in our attack.

\subsection{Cache Attacks}

The cache attack is the most well-known representative of the general
class of micro-architectural attacks, which are defined by Acii�mez
in his excellent survey \cite{DBLP:conf/ccs/Aciicmez07} as attacks
which ``exploit deeper processor ingredients below the trust architecture
boundary'' to recover secrets from various secure systems. Cache
attacks make use of the fact that, regardless of higher-level security
mechanisms such as sandboxing, virtual memory, privilege rings, hypervisors
etc., both secure and insecure processes can interact through their
shared use of the cache. This allows an attacker to craft a ``spy''
process which can measure and make inferences about the internal state
of a secure process through their shared use of the cache. First identified
by Hu in 1992 \cite{DBLP:conf/sp/Hu92} , several results have shown
how the cache side-channel can be used to recover AES keys \cite{DBLP:conf/ctrsa/OsvikST06,djb-cache},
RSA keys \cite{2005-percival-cache}, and even allow one virtual machine
to compromise another virtual machine running on the same host \cite{DBLP:conf/ccs/RistenpartTSS09}.

Our attack is modeled after the \textsc{Prime+Probe }attack method,
first described by Osvik et al. in \cite{DBLP:conf/ctrsa/OsvikST06}
in the context of the L1 cache. The attack was later extended by Yarom
et al. in \cite{Yarom_LGHL_15} to last-level caches on systems with
large pages enabled, and we extend it in this work to last-level caches
in the more common case of 4K-sized pages. In general, the \textsc{Prime+Probe}
attack follows a four-step pattern. In the first step, the attacker
creates one or more \textbf{eviction sets}. An eviction set is a set
of locations in memory which, when accessed, can take over a single
cache line which is also used by the victim process. In the second
step, the attacker \textbf{primes} the cache set by accessing the
eviction set. This forces the eviction of the victim's code or instructions
from the cache set and brings it to a known state. In the third step,
the attacker triggers or simply waits for the victim to execute and
potentially utilize the cache. Finally, the attacker \textbf{probes}
the cache set by accessing the eviction set yet again. A low access
latency suggests that the attacker's code or data is still in the
cache, while a higher access latency suggests that the victim's code
made use of the cache set, thereby teaching the attacker about the
victim's internal state. The actual timing measurement is carried
out by using the unprivileged assembler instruction \texttt{\textsc{rdtsc}},
which provides a very sensitive measurement of the processor's cycle
count. Iterating over the linked list also serves a secondary purpose
by forcing the cache set yet again into an attacker-controlled state,
thus preparing for the next round of measurements.

\subsection{The Web Runtime Environment}

Javascript is a dynamically typed, object-based scripting language
with runtime evaluation, which powers the client side of the modern
web. Javascript code is delivered to the browser runtime in source-code
form and is compiled and optimized by the browser using a just-in-time
mechanism. The fierce competition between different browser vendors
resulted in an intense focus on improving Javascript performance.
As a result, Javascript code performs in some scenarios on a level
which is on par with that of native code. 

The core functionality of the Javascript language is defined by the
ECMA industry association in Standard ECMA-262 \cite{ecma-262}. The
language standard is complemented by a large set of application programming
interfaces (APIs) defined by the World Wide Web Consortium \cite{javascript-api},
which make the language practical for developing web content. The
Javascript API set is constantly evolving, and browser vendors add
support to new APIs over time according to their own development schedules.
Two specific APIs which are of use to us in this work are the Typed
Array Specification \cite{TypedArray}, which allows efficient access
to unstructured binary data, and the High Resolution Time API \cite{w3c-time},
which provides sub-millisecond time measurements to Javascript programs.
As we show in Subsection \ref{sub:Is-Your-Computer}, a large majority
of Web browsers in use today support both of these APIs.

Javascript code runs in a highly \textbf{sandboxed} environment --
code delivered via Javascript has highly restricted access to the
system. For example, Javascript code cannot open files, even for reading,
without the permission of the user. Javascript code cannot execute
native language code or load native code libraries. Most significantly,
Javascript code has \textbf{no notion of pointers}. Thus, it is impossible
to determine even the virtual address of a Javascript variable.

\subsection{Our Contribution}

Our objective was to craft a last-level cache attack which can be
deployed over the web. This process is quite challenging since Javascript
code cannot load shared libraries or execute native language programs,
and since Javascript code is forced to make timing measurements using
scripting language function calls instead of using dedicated assembler
instruction calls. These challenges notwithstanding, we have been
able to successfully extend cache attacks to the web-based environment:
\begin{itemize}
\item We present a novel method of creating a \textbf{non-canonical eviction
set} for the last-level cache. In contrast to \cite{Yarom_LGHL_15},
our method does not require the system to be configured for large
page support, and as such can immediately be applied to a wider variety
of desktop and server systems. We show that our method runs in a practical
time even when implemented in Javascript.
\item We present a \textbf{fully functional last-level cache attack} \textbf{using
unprivileged Javascript code}. We evaluate its performance using a
covert channel method, both between different processes running on
the same machine and between a VM client and its host. The nominal
capacity of the Javascript-based channel is on the order of hundreds
of kilobits per second, comparable to that of the native code approach
of \cite{Yarom_LGHL_15}.
\item We show how cache-based methods can be used to effectively \textbf{track
the behavior} of the user. This application of cache attacks is more
relevant to our attack model than the cryptanalytic applications often
explored in other works.
\item Finally, we \textbf{describe possible countermeasures} to our attack
and discuss their systemwide cost.
\end{itemize}
\textbf{Document Structure: }In Section \ref{sec:Attack-Methodology}
we presents the design and implementation of the different steps of
our attack methodology. In Section \ref{sec:Cache-Based-Covert} we
present a covert channel constructed using our attack methodology
and evaluate its performance. In Section \ref{sec:User-Behaviour-Tracking}
we investigate the use of cache-based attacks for tracking user behavior
both inside and outside the browser. Finally, Section \ref{sec:Discussion}
concludes the paper with a discussion of countermeasures and open
research challenges.

\section{\label{sec:Attack-Methodology}Attack Methodology}

As described in the previous section, the four steps involved in a
successful \textsc{Prime+Probe }attack are: creating an eviction set
for one or more relevant cache sets, priming the cache set, triggering
the victim operation and finally probing the cache set again. While
the actual priming and probing are pretty straightforward to implement,
finding cache sets which correlate to interesting system behaviors
and creating eviction sets for them is less trivial. In this Section
we describe how each of these steps was implemented in Javascript.

\subsection{\label{sub:Creating-an-Eviction}Creating an Eviction Set}

\subsubsection{Design}

As stated in \cite{Yarom_LGHL_15}, the first step of a \textsc{Prime+Probe
}attack is to create an eviction set for a certain desired cache set
shared with a victim process. This eviction set consists of a set
of variables which are all mapped by the CPU into the same cache set.
The use of a linked list is meant to defeat the CPU's memory prefetching
and pipelining optimizations, as suggested by \cite{DBLP:conf/ccs/RistenpartTSS09}.
We first show how we create an eviction set for an arbitrary cache
set, and later address the problem of finding which cache set is shared
with the victim.

As discussed in \cite{DBLP:conf/ctrsa/OsvikST06}, the L1 cache determines
the set assignment for a variable based the lower bits of its virtual
address. Since the attacker is assumed to know the virtual addresses
of its own variables, it was thus straightforward to create an eviction
set in the L1 attack model. In contrast, set assignments for variables
in the LLC are made by reference to their physical memory address,
which are not generally available to an unprivileged process. The
authors of \cite{Yarom_LGHL_15} partially circumvented this problem
by assuming that the system is operating in large page mode, in which
the lower 21 bits of the physical and virtual addresses are identical,
and by the additional use of an iterative algorithm to resolve the
unknown upper (slice) bits of the cache set index.

In the attack model we consider, the system is running in the traditional
4K page mode, where only the lower 12 bits of the physical and virtual
addresses are identical. To our further difficulty, Javascript has
no notion of pointers, so even the virtual addresses of our own variables
are unknown to us. 

The mapping of 64-bit physical memory addresses into 13-bit cache
set indices was investigated by Hund et al. in \cite{DBLP:conf/sp/HundWH13}.
They discovered that accessing a contiguous 8MB ``eviction buffer''
of physical memory will completely invalidate all cache sets in the
L3 cache. While we could not allocate such an eviction buffer in user-mode
(indeed, the work of \cite{DBLP:conf/sp/HundWH13} was assisted by
a kernel-mode driver), we allocated an 8MB byte array in virtual memory
using Javascript (which was assigned by the operating system into
an arbitrary and non-contiguous set of 4K physical memory pages),
and measured the system-wide effects of iterating over this buffer.
We discovered that access latencies to unrelated variables in memory
were slowed down by a noticeable amount when we accessed them immediately
after iterating through this eviction buffer. We also discovered that
the slowdown effect persisted even if we did not access the entire
buffer, but rather accessed it in offsets of once per every 64 bytes.
However, it was not immediately clear how to map each of the 131K
offsets we accessed inside this eviction buffer into each of the 8192
possible cache sets, since we did not know the physical memory locations
of the various pages of our buffer.

A naive approach to solving this problem would be to fix an arbitrary
``victim'' address in memory, then find by brute force which set
of 12 out of the 131K offsets share a set with this address. To do
so, we could fix some subset of the 131K offsets, then measure whether
the access latency to this victim address is increased after iterating
through these offsets. If the latency increases, this means the subset
contains the 12 addresses sharing the set with the victim address.
If the latency does not change, then the subset does not contain at
least one of these 12 addresses, allowing the victim address to remain
in the cache. By repeating this process 8192 times, each time with
a different victim address, we would be able to identify each cache
set and create our data structure.

An immediate application of this heuristic would take an impractically
long time to run. Fortunately, the page frame size of the Intel MMU,
as described in Subsection \ref{sub:The-Memory-Architecture}, could
be used to our great advantage. Since virtual memory is page aligned,
the lower 12 bits of each virtual memory address are identical to
the lower 12 bits of each physical memory address. According to Hund
et al., 6 of these 12 bits are used in uniquely determining the cache
set index. Thus, an offset in our eviction buffer cannot be the same
cache set as all 131K other offsets, but rather only with the 8K other
offsets sharing address bits 12 to 6. In addition, discovering a single
cache set can immediately teach us about 63 additional cache sets
located in the same page frame. Joined with the discovery that Javascript
allocates large data buffers along page frame boundaries, this led
to the greedy algorithm described in Algorithm \ref{alg:Identifying-a-cache}.

\begin{algorithm}
Let $S$ be the set of unmapped pages, and address $x$ be an arbitrary
page-aligned address in memory
\begin{enumerate}
\item Repeat $k$ times:

\begin{enumerate}
\item Iteratively access all members of $S$
\item Measure $t_{1}$, the time it takes to access $x$
\item Select a random page $s$ from $S$ and remove it
\item Iteratively access all members of $S\backslash s$
\item Measure $t_{2}$, the time it takes to access $x$
\item If removing page $s$ caused the memory access to speed up considerably
(i.e., $t_{1}-t_{2}>thres$), then this page is part of the same set
as $x$. Place it back into $S$.
\item If removing page $s$ did not cause memory access to speed up considerably,
then this address is not part of the same set as $x$.
\end{enumerate}
\item If $\left|S\right|=12$, return $S$. Otherwise report failure.
\end{enumerate}
\protect\caption{\label{alg:Identifying-a-cache}Profiling a cache set}
\end{algorithm}

By running Algorithm \ref{alg:Identifying-a-cache} multiple times,
we can gradually create eviction sets covering most of the cache,
except for those parts which are accessed by the Javascript runtime
itself. We note that, in contrast to the eviction sets created by
the algorithm of \cite{Yarom_LGHL_15}, our eviction set is \textbf{non-canonical}
-- since Javascript has no notion of pointers, we cannot identify
which of the CPU's cache sets corresponds to any particular eviction
set we discover. Furthermore, running the algorithm multiple times
on the same system will result in a different mapping each time it
is run. This property stems from the use of traditional 4K pages instead
of large 2MB pages, and will hold even if the eviction sets are created
using native code and not Javascript.

\subsubsection{Evaluation}

We implemented Algorithm \ref{alg:Identifying-a-cache} in Javascript
and evaluated it on Intel machines using CPUs from the Ivy Bridge,
Sandy Bridge and Haswell families, running the latest versions of
Safari and Firefox on Mac OS Yosemite and Ubuntu 14.04 LTS, respectively.
The systems were not configured to use large pages, but instead were
running with the default 4K page size. The code snippet shown in Listing
\ref{lis:Javascript-code-to} shows lines 1.d and 1.e of the algorithm,
and demonstrate how we iterate over the linked list and measure latencies
using Javascript. The algorithm requires some additional steps to
run under Chrome and under Internet Explorer, which we describe in
Subsection \ref{sub:Is-Your-Computer}.

\begin{lstlisting}[caption={Javascript code to invalidate a cache
set, then measure access time},label={lis:Javascript-code-to},basicstyle={\ttfamily},captionpos=b,float,language=Java]
  // Invalidate the cache set
  var currentEntry = startAddress;
  do {
	currentEntry = 
      probeView.getUint32(currentEntry);
  } while (currentEntry != startAddress);

  // Measure access time
  var startTime = 
    window.performance.now();
  currentEntry = 
    primeView.getUint32(variableToAccess); 
  var endTime = window.performance.now();
\end{lstlisting}

\begin{figure}

\begin{centering}
\includegraphics[width=0.9\columnwidth]{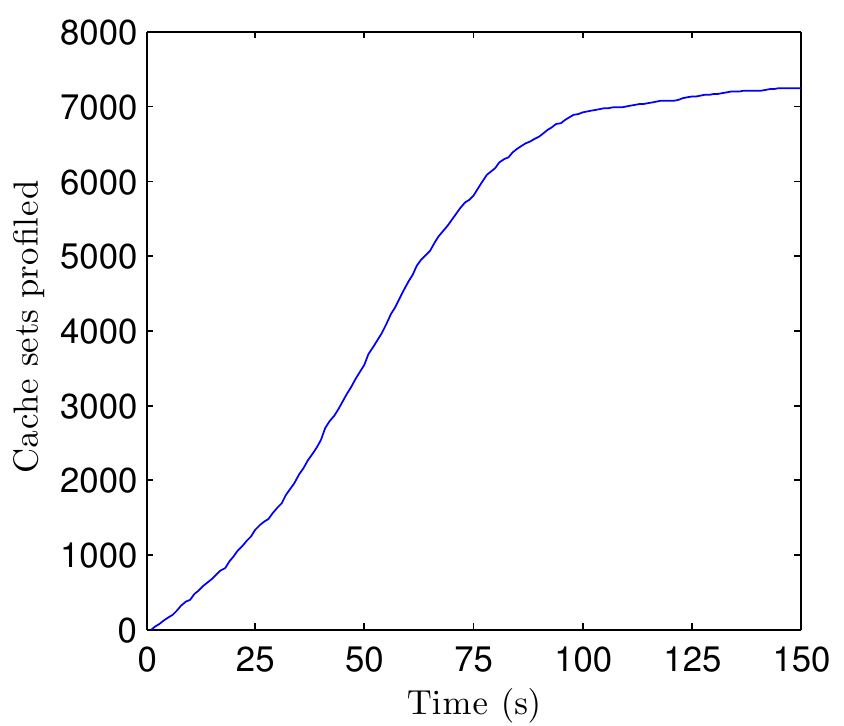}
\par\end{centering}

\protect\caption{\label{fig:Cumulative-performance-of}Cumulative performance of the
profiling algorithm}

\end{figure}

Figure \ref{fig:Cumulative-performance-of} shows the performance
of the profiling algorithm, as evaluated on an Intel i7-3720QM running
Firefox 35.0.1 for Mac OS 10.10.2. We were pleased to find that the
algorithm was able to map more than 25\% of the cache in under 30
seconds of operation, and more than 50\% of the cache after 1 minute.
The algorithm seems very simple to parallelize, since most of the
execution time is spent on data structure maintenance and only a minority
of it is actually spent in the actual invalidate-and-measure portion.
The entire algorithm fits into less than 500 lines of Javascript code.

\begin{figure}
\begin{centering}
\includegraphics[width=0.9\columnwidth]{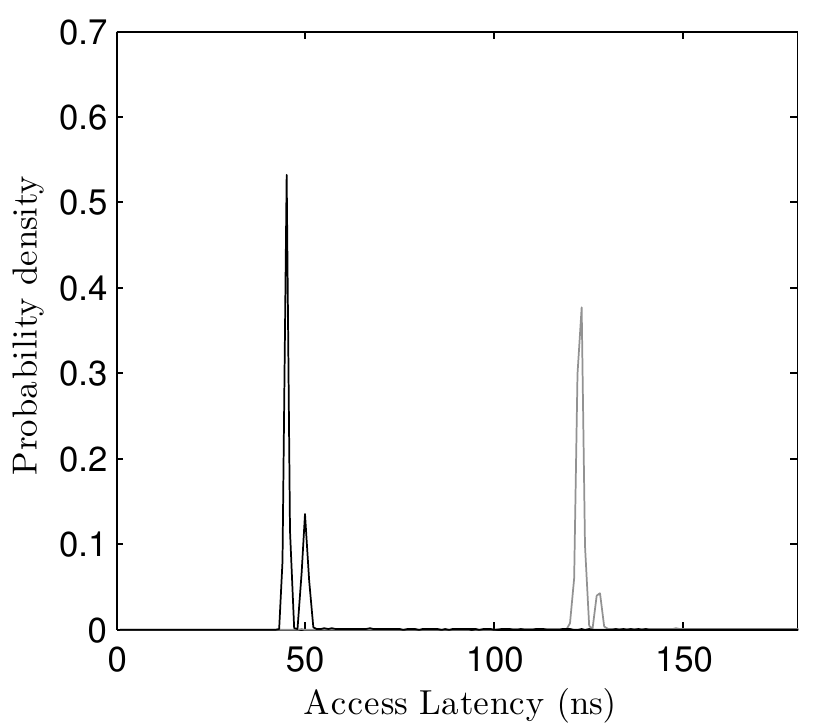}
\par\end{centering}

\protect\caption{\label{fig:Access-times}Probability distribution of access times
for flushed vs. un-flushed variable (Haswell CPU)}
\end{figure}

\begin{figure}
\begin{centering}
\includegraphics[width=0.9\columnwidth]{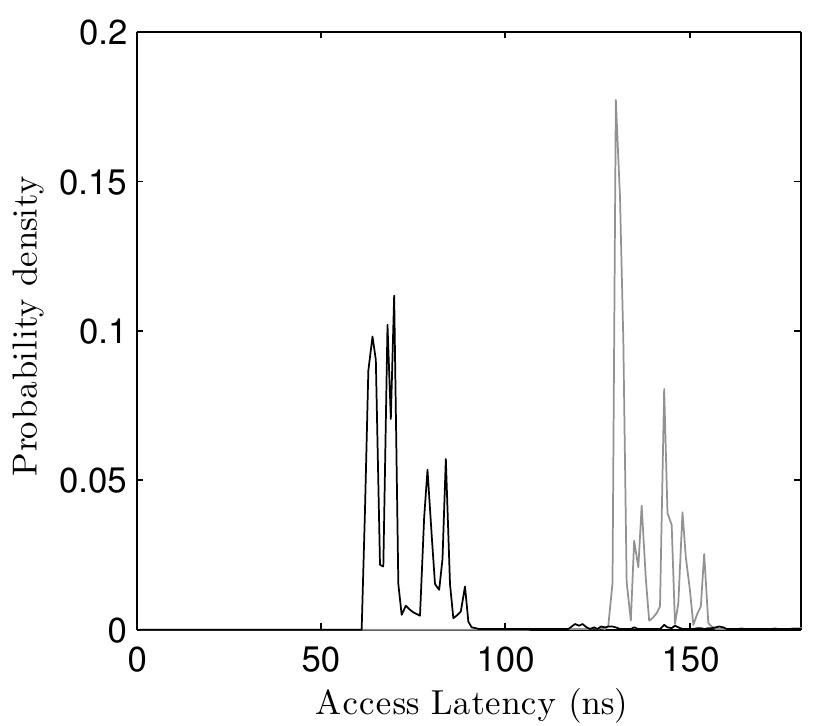}
\par\end{centering}

\protect\caption{\label{fig:Access-times-1}Probability distribution of access times
for flushed vs. un-flushed variable (Sandy Bridge CPU)}
\end{figure}

To verify that our algorithm was indeed capable of identifying cache
sets, we designed an experiment that compares the access latencies
for a flushed and an un-flushed variable. Figure \ref{fig:Access-times}
shows two probability distribution functions comparing of the time
required to access a variable which has recently been flushed from
the cache using our method (gray line) with the time required to access
a variable which currently resides in the cache set (black line).
The timing measurements were carried out using Javascript's high resolution
timer, and thus include the additional delay imposed by the Javascript
runtime. It is clear to see that the two distributions are distinguishable,
confirming the correct operation of our profiling method. Figure \ref{fig:Access-times-1}
shows a similar plot captured on an older-generation Sandy Bridge
CPU, which includes 16 entries per cache set.

By selecting a group of cache sets and repeatedly measuring their
access latencies over time, the attacker is provided with a very detailed
picture of the real-time activity of the cache. We call the visual
representation of this image a ``memorygram'', since it is looks
quite similar to an audio spectrogram. 

\begin{figure*}
\begin{centering}
\includegraphics[width=0.9\textwidth]{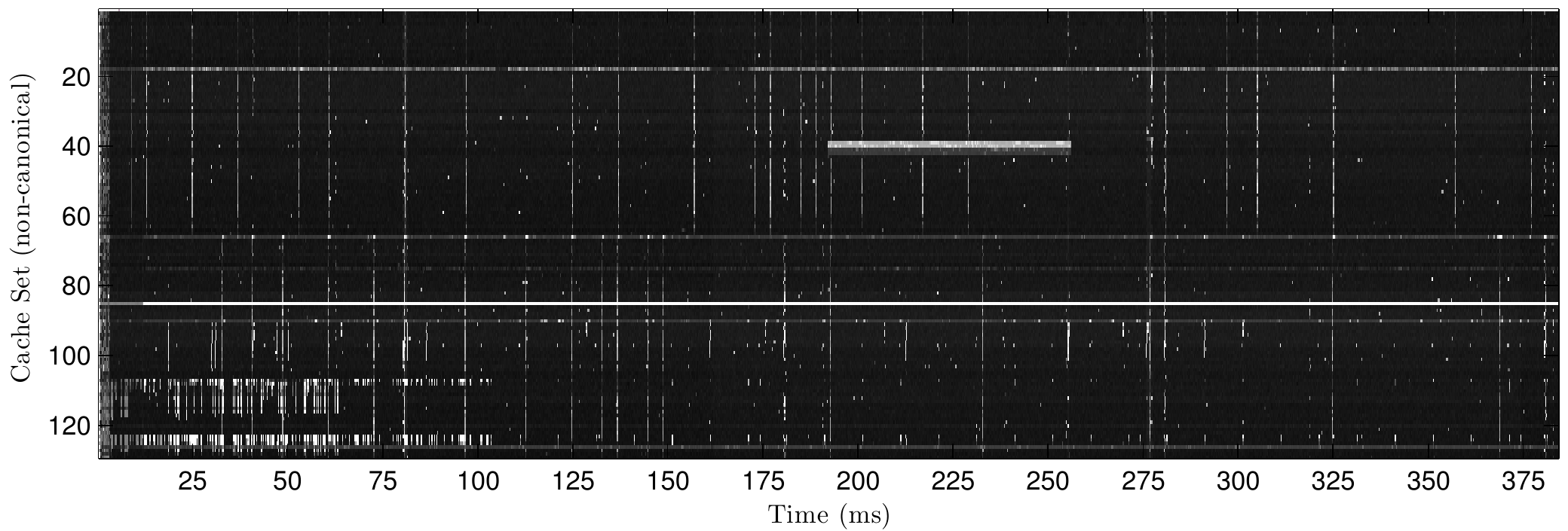}
\par\end{centering}

\protect\caption{\label{fig:Sample-memorygram}Sample memorygram}
\end{figure*}

A sample memorygram, collected over an idle period of 400ms, is presented
in Figure \ref{fig:Sample-memorygram}. The X axis corresponds to
time, while the Y axis corresponds to different cache sets. The sample
shown has a temporal resolution of $250\text{\textmu s}$ and monitors
total of 128 cache sets. The intensity of each pixel corresponds to
the access latency of this particular cache set at this particular
time, with black representing a low latency, indicating no other process
accessed this cache set between the previous measurement and this
one, and white representing a higher latency, suggesting that the
attacker's data was evicted from the cache between this measurement
and the previous one.

Observing this memorygram can provide several immediate insights.
First, it is clear to see that despite the use of Javascript timers
instead of machine language instructions, measurement jitter is quite
low active and inactive sets are clearly differentiated. It is also
easy to notice several vertical line segments in the memorygram, indicating
multiple adjacent cache sets which were all active during the same
time period. Since consecutive cache sets (within the same page frame)
correspond to consecutive addresses in physical memory, we believe
this signal indicates the execution of a function call which spans
more than 64 bytes of assembler instructions. Several smaller groups
of cache sets are also accessed together. We theorize that the these
smaller groups correspond to variable accesses. Finally, the white
horizontal line indicates a variable which is constantly accessed
during our measurements. This variable probably belongs to the measurement
code or to the underlying Javascript runtime. It is remarkable that
such a wealth of information about the system is available to an unprivileged
webpage!

\subsection{Identifying Interesting Regions in the Cache}

The eviction set allows the attacker to monitor the activity of arbitrary
sets of the cache. Since the eviction set we receive is non-canonical,
the attacker must now correlate the cache sets he has profiled to
data or code locations belonging to the victim. This learning/classification
problem was addressed earlier by Zhang et al. in \cite{DBLP:conf/ccs/ZhangJRR12}
and by Yarom et al. in \cite{Yarom_LGHL_15}, where various machine
learning methods such as SVM were used to derive meaning from the
output of cache latency measurements. 

To effectively carry out the learning step, the attacker needs to
induce the victim to perform an action, then examine which cache sets
were touched by this action, as formally defined in Algorithm \ref{alg:set-detection}. 

\begin{algorithm}
Let $S_{i}$ be the data structure matched to eviction set $i$
\begin{enumerate}
\item For each set $i$:

\begin{enumerate}
\item Iteratively access all members of $S_{i}$ to prime the cache set
\item Measure the time it takes to iteratively access all members of $S_{i}$ 
\item Perform an interesting operation
\item Measure once more the time it takes to iteratively access all members
of $S_{i}$ 
\item If performing the interesting operation caused the access time to
slow down considerably, then the operation was associated with cache
set $i$.
\end{enumerate}
\end{enumerate}
\protect\caption{\label{alg:set-detection}Interesting Regions in the Cache}
\end{algorithm}

Finding a function for step (c) of the algorithm was actually quite
challenging due to the limited permissions granted to Javascript code.
This can be contrasted with the ability of Apecechea et al. to trigger
a minimal kernel operation by invoking an empty \texttt{sysenter}
call \cite{DBLP:conf/raid/ApececheaIES14}. To carry out this step,
we had to survey the Javascript runtime to discover function calls
which may trigger interesting behavior, such as file access, network
access, memory allocation, etc. We were also interested in functions
which take a relatively short time to run and left no background ``tails''
such as garbage collection which would impact our measurement in step
(d). Several such functions were discovered in a different context
by Ho et al. in \cite{DBLP:conf/woot/HoBBP14}. Another approach would
be to induce the user to perform an interesting behavior (such as
pressing a key on his keyboard) on the behalf of the attacker. The
learning process in this case might be structured (where the attacker
knows exactly when the victim operation was executed), or unstructured
(where the attacker can only assume that relatively busy periods of
system activity are due to victim operations. We make use of both
of these approaches in the attack we present in Section \ref{sec:User-Behaviour-Tracking}. 

Since our code will always detect activity caused by the Javascript
runtime, the high performance timer code, and other components of
the web browser which are running regardless of the call being executed,
we actually called two similar functions and examined the \textbf{difference}
between the activity profile of the two evaluations to identify relevant
cache sets.

\section{\label{sec:Cache-Based-Covert}A Cache-Based Covert Channel in Javascript}

\subsection{Motivation}

As shown in \cite{Yarom_LGHL_15}, last-level cache access patterns
can be used to construct a high-bandwidth covert channel and effectively
exfiltrate sensitive information between virtual machines co-resident
on the same physical host. In our particular attack model, in which
the attacker is not in a co-resident virtual machine but rather inside
a webpage, the motivation for a covert channel is different but still
very interesting.

By way of motivation, let us assume that a Security Agency is tracking
the criminal mastermind Bob. Making use of a spear phishing campaign,
the Agency installs a piece of software of its own choosing, commonly
referred to as an Advanced Persistent Threat (APT), on Bob's personal
computer. The APT is designed to log incriminating information about
Bob and send it to the Agency's secret servers. Bob is however highly
security-savvy, and is using an operation system which enforces strict
Information Flow Tracking \cite{DBLP:conf/osdi/ZeldovichBKM06}. This
operating system feature prevents the APT from accessing the network
after it accesses any file containing private user data.

Javascript-based cache attacks can immediately be put to use to allow
the Agency to operate in such a scenario, as long as Bob can be enticed
to view a website controlled by the Security Agency. Instead of transmitting
the private user data over the network, the APT will use the cache
side-channel to communicate with the malicious website, without setting
off the flow tracking capabilities of the operating system.

This case study is inspired by the ``RF retro-reflector'' design
attributed to a certain Security Agency, in which a collection device
such as a microphone does not transmit the collected signal directly,
but instead modulates the collected signal onto an ``illuminating
signal'' sent to it by an external ``collection device''.

\subsubsection{Design}

The design of our covert channel system was influenced by two requirements:
first, we wanted the transmitter part to be as simple as possible,
and in particular we did not want it to carry out the eviction set
algorithm of Subsection \ref{sub:Creating-an-Eviction}. Second, since
the receiver's eviction set is non-canonical, it should be as simple
as possible for the receiver to search for the sets onto which the
transmitter was modulating its signal. 

To satisfy these requirements, our transmitter/APT simply allocates
a 4K array in its own memory and continuously modulates the collected
data into the pattern of memory accesses to this array. There are
64 cache sets covered by this 4K array, allowing the APT to transmit
64 bits per time period. To make sure the memory accesses are easily
located by the receiver, the same access pattern is repeated in several
additional copies of the array. Thus, a considerable percentage of
the cache is actually exercised by the transmitter, in contrast to
the method of \cite{Yarom_LGHL_15} which assumes a canonical eviction
set, and thus only activates two lines.

The receiver code profiles the system's physical memory, then searches
for one of the page frames containing the data modulated by the APT.
The data can then be demodulated from the memory access pattern and
uploaded back to the server, all without violating the information
flow tracking protections.

\subsubsection{Evaluation}

Our attacker model assumes that the transmitter part is written in
(relatively fast) native language, while the receiver part is implemented
in Javascript. Thus, we assumed that the limiting factor in the performance
of our system is the sampling speed of the malicious website. 

To evaluate the bandwidth of this covert channel, we wrote a simple
program that iterates over memory in a predetermined pattern (in our
case, a bitmap containing the word ``Usenix''). Next, we attempted
to search for this memory access pattern using a Javascript cache
attack, then measured the maximum sampling frequency at which the
Javascript code could be run.

\begin{figure}
\begin{centering}
\includegraphics[width=0.9\columnwidth]{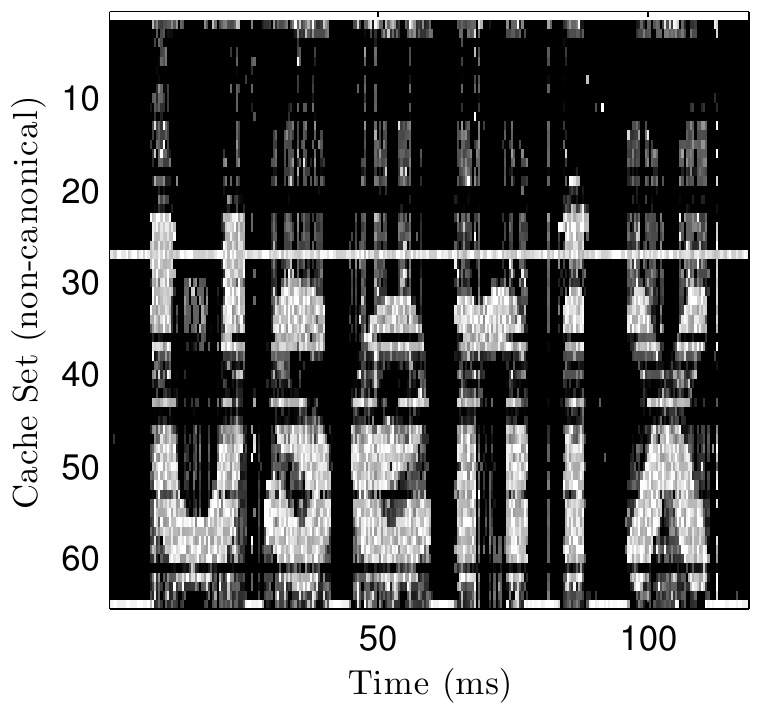}
\par\end{centering}

\protect\caption{\label{fig:covert-host-to-host}A host-to-host covert channel}
\end{figure}

Figure \ref{fig:covert-host-to-host} shows a memorygram capturing
an execution of this covert channel. The nominal bandwidth of the
covert channel was measured to be approximately 320kbps, a figure
which compares well with the 1.2Mbps bandwidth achieved by the native
code cross-VM covert channel implemented by \cite{Yarom_LGHL_15}.

\begin{figure}
\begin{centering}
\includegraphics[width=0.9\columnwidth]{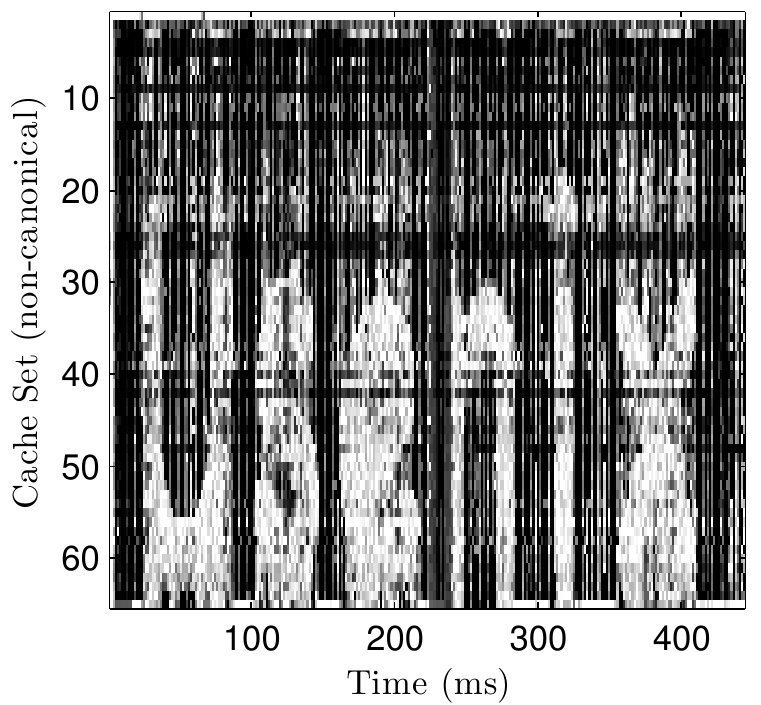}
\par\end{centering}

\protect\caption{\label{fig:covert-host-to-vm}A host-to-VM covert channel}
\end{figure}

Figure \ref{fig:covert-host-to-vm} shows a similar memorygram where
the receiver code is not running directly on the host, but rather
on a virtual machine (Firefox 34 running on Ubuntu 14.01 inside VMWare
Fusion 7.1.0). While the peak bandwidth of the in this scenario was
severely degraded to approximately 8kbps, the fact that a webpage
running inside a virtual machine is capable of probing the underlying
hardware is still quite surprising.

\section{\label{sec:User-Behaviour-Tracking}User Behavior Tracking Through
Cache Attacks}

Most works which evaluate cache attacks assume that the attacker and
the victim share a colocated machine inside a cloud-provider data
center. Such a machine is not typically configured to accept interactive
input, and accordingly most works in this field focus on the recovery
of cryptographic keys or other secret state elements, such as random
number generator states \cite{DBLP:conf/ccs/ZhangJRR14}. For this
work, we chose to examine how cache attacks can be used to track the
interactive behavior of the user, a threat which is more relevant
to the attack model we consider. We note that \cite{DBLP:conf/ccs/RistenpartTSS09}
have already attempted to track keystroke timing events using coarse-grained
measurements of system load on the L1 cache.

This case study shows how a malicious webpage can track a user's activity
using a cache attack. In the attack presented below, we assume that
the user has loaded a malicious webpage in a background tab or window,
and is carrying out sensitive operations in another tab, or even in
a completely different application with no Internet connectivity.

We chose to focus on mouse and network activity because the operating
system code that handles them is non-negligible. Thus, we expected
them to have a relatively large cache footprint. They are also easily
triggered by content running within the restricted Javascript security
model, as we describe below.

\subsection{Design}

The structure of both attacks is similar. First, the profiling phase
is carried out, allowing the attacker to probe individual cache sets
using Javascript. Next, during a training phase, the activity to be
detected (i.e. network activity or mouse activity) is triggered, and
the cache activity is sampled multiple times with a very high temporal
resolution. While the network activity was triggered directly by the
measurement script (by executing a network request), we simply waved
the mouse around over the webpage during the training period %
\footnote{In a full attack, the user can be enticed to move the mouse by having
him play a game or fill out a form. %
}. 

By comparing the cache activity during the idle and active periods
of the training phase, the attacker learns which cache sets are uniquely
active during the relevant activity and trains a classifier on these
cache sets. Finally, during the classification phase, the attacker
monitors the interesting cache sets over time to learn about the user's
activity. 

We used a basic unstructured training process, assuming that the most
intensive operation performed by the system during the training phase
would be the one being measured. To take advantage of this property,
we calculated the Hamming weight of each measurement over time (equivalent
to the count of cache sets which are active during a certain time
period), then applied a k-means clustering of these Hamming weights
to divide the measurements into several clusters. We then calculated
the mean access latency of each cache set in every cluster, arriving
at a \emph{centroid }for each cluster. To classify an unknown measurement
vector, we measured the Euclidean distance between this vector and
each of these centroids, classifying it as the closest one. 

In the classification phase, we generated network traffic using the
command-line tool wget and moved the mouse outside of the browser
window.  To provide ground truth for the network activity scenario,
we concurrently measured the traffic on the system using tcpdump,
then mapped the timestamps logged by tcpdump to the times detected
by our classifier. To provide ground truth for the mouse activity
scenario, we wrote a webpage that timestamps and logs all mouse events,
then moved the mouse over this webpage. We stress that the mouse-logging
webpage was run on a different browser (Chrome) than the measuring
code (Firefox).

\subsection{Evaluation}

\begin{figure}
\begin{centering}
\includegraphics[width=0.9\columnwidth]{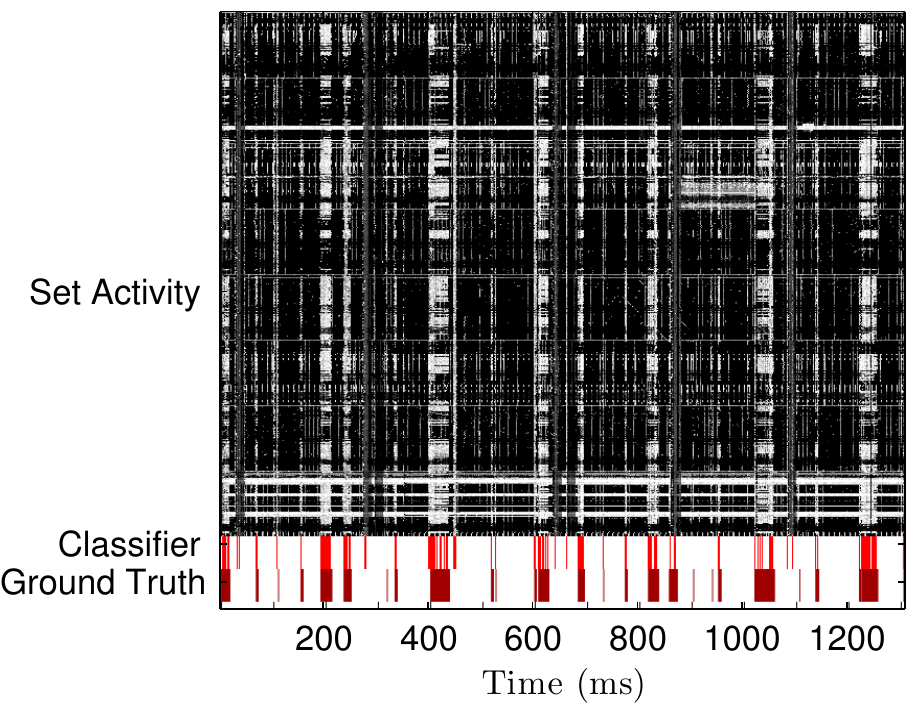}
\par\end{centering}

\protect\caption{\label{fig:network-activity}Network activity detection}
\end{figure}

\begin{figure}
\begin{centering}
\includegraphics[width=0.9\columnwidth]{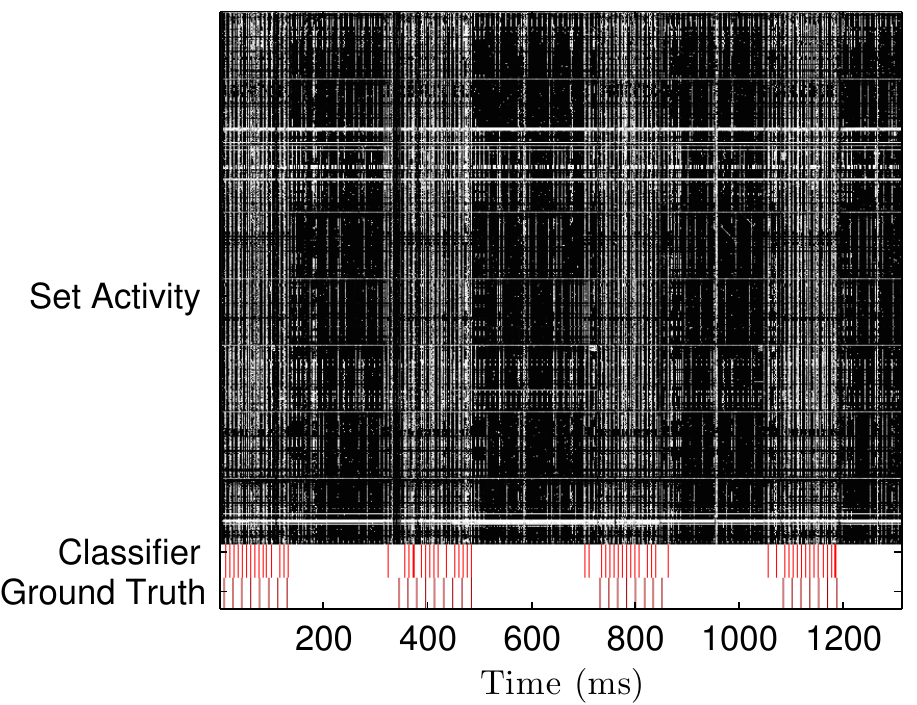}
\par\end{centering}

\protect\caption{\label{fig:mouse-activity}Mouse activity detection}
\end{figure}

The results of the activity measurement are shown in Figures \ref{fig:network-activity}
and \ref{fig:mouse-activity}. The top part of both figures shows
the real-time activity of a subset of the cache. On the bottom part
of each figure are the classifier outputs, together with the ground
truth which was collected externally. As the Figures show, our extremely
simple classifier was quite capable of detecting mouse and network
activity. The performance of the attack can be improved without a
doubt by using more advanced training and classification techniques.
We stress that the mouse activity detector did not detect network
activity, and vice versa.

The classifier's measurement rate was only 500Hz. As a result, it
could not count individual packets but rather periods of network activity
and inactivity. In contrast, our mouse detection code actually logged
more events than the ground truth collection code. This is due to
the fact that the Chrome browser throttles mouse events to web pages
down to a rate of approximately 60Hz.

Detecting network activity can be a stepping stone toward a deeper
insight of the user's activity, as famously demonstrated by Chen et
al. in \cite{DBLP:conf/sp/ChenWWZ10}. In essence, while Chen et al.
assumed a network-level attacker which can monitor all incoming and
outgoing traffic to the victim, the techniques presented here can
enable any malicious website to monitor the concurrent web activities
of its users. The attack can be bolstered by more indicators, such
as memory allocations (as explored by \cite{DBLP:conf/sp/JanaS12a}),
DOM layout events, disk writes and so on.

\section{\label{sec:Discussion}Discussion}

This work shows that side-channel attacks have a much wider reach
than previously expected. Instead of being relevant only for very
specific attacker scenarios, the attack proposed here can be mounted
against most computers connected to the Internet. The fact that so
many systems are suddenly vulnerable to side-channel attacks suggests
that side-channel resistant algorithms and systems should be the norm,
rather than the exception.

\subsection{\label{sub:Is-Your-Computer}Prevalence of Vulnerable Systems}

Our attack requires a personal computer powered by an Intel CPU based
on the Sandy Bridge, Ivy Bridge, Haswell or Broadwell micro-architectures.
According to data from IDC, more than 80\% of all PCs sold after 2011
satisfy this requirement. We furthermore assume that the user is using
a web browser which supports the HTML 5 High Resolution Time API and
the Typed Arrays specification. Table \ref{tab:Prevalence-of-vulnerable}
notes the earliest version at which these APIs are supported for each
of the common browser brands, as well as the proportion of global
Internet traffic coming from vulnerable browser versions, according
to StatCounter GlobalStats measurements as of January 2015 \cite{StatCounter}.
As the table shows, more than 80\% of desktop browsers in use today
 are vulnerable to the attack we describe.

\begin{table*}
\begin{centering}
\begin{tabular}{|c|c|c|c|}
\hline 
Browser brand & High Resolution Time Support & Typed Arrays Support & Worldwide prevalence\tabularnewline
\hline 
\hline 
Internet Explorer & 10 & 11 & 11.77\%\tabularnewline
\hline 
Safari & 8 & 6 & 1.86\%\tabularnewline
\hline 
Chrome & 20%
\footnote{The resolution of this timer was lower than expected on Windows versions
20 to 38, due to a bug) %
} & 7 & 50.53\%\tabularnewline
\hline 
Firefox & 15 & 4 & 17.67\%\tabularnewline
\hline 
Opera & 15 & 12.1 & 1.2\%\tabularnewline
\hline 
Total & -- & -- & 83.03\%\tabularnewline
\hline 
\end{tabular}
\par\end{centering}

\protect\caption{\label{tab:Prevalence-of-vulnerable}Prevalence of vulnerable desktop
browsers, according to \cite{StatCounter}}
\end{table*}

The effectiveness of our attack depends on being able to perform precise
measurements using the Javascript High Resolution Time API. While
the W3C recommendation of this API \cite{w3c-time} specifies that
the a high-resolution timestamp should be ``a number of milliseconds
accurate to a thousandth of a millisecond'', the maximum resolution
of this value is not specified, and indeed varies between browser
versions and operating systems. In our testing we discovered, for
instance, that the actual resolution of this timestamp for Safari
for MacOS was on the order of nanoseconds, while Internet Explorer
for Windows had a 0.8\textmu s resolution. Chrome, on the other hand,
offered a uniform resolution of 1\textmu{} on all operating systems
we tested. 

Since, as shown in Figure \ref{fig:Access-times}, the timing difference
between a single cache hit and a single cache miss is on the order
of 50ns, the profiling and measurement algorithms need to be slightly
modified to support systems with coarser-grained timing resolution.
In the profiling stage, instead of measuring a single cache miss we
repeat the memory access cycle multiple times to amplify the time
difference. For the measurement stage, we cannot amplify a single
cache miss, but we can take advantage of the fact that code access
typically invalidates multiple consecutive cache sets from the same
page frame. As long as at least 20 out of the 64 cache sets in a single
page frame register a cache miss, our attack is successful even with
microsecond time resolution. 

The attack we propose is also easily applied to mobile devices such
as smartphones and tablets. It should be noted that the Android Browser
supports High Resolution Time and Typed Arrays starting from version
4.4, but at the time of writing the most recent version of iOS Safari
(8.1) did not support the High Resolution Time API.

\subsection{Countermeasures}

The attacks described in this report are possible because of a confluence
of design and implementation decisions starting at the micro-architectural
level and ending at the Javascript runtime: The method of mapping
a physical memory address to cache set; the inclusive cache micro-architecture;
Javascript's high-speed memory access and high-resolution timer; and
finally, Javascript's permission model. Mitigation steps can be applied
at each of these junctions, but each will impose a drawback on the
benign uses of the system.

On the \textbf{micro-architectural} level, changes to the way physical
memory addresses are mapped to cache lines will severely confound
our attack, which makes great use the fact that 6 of the lower 12
bits of the address are used directly to select a cache set. Similarly,
the move to an exclusive cache micro-architecture, instead of an inclusive
one, will make it impossible for our code to trivially evict entries
from the L1 cache, making measurement much more difficult. These two
design decisions, however, were chosen deliberately to make the CPU
more efficient in its design and in its use of cache memory, and changing
them will exact a performance cost on many other applications. In
addition, modifying a CPU's micro-architecture is far from trivial,
and definitely impossible as an upgrade to already deployed hardware.

On the \textbf{Javascript} level, it seems that somewhat reducing
the resolution of the high-resolution timer will make this attack
more difficult to launch. However, the high-resolution timer was created
to address a real need of Javascript developers for applications ranging
from music and games to augmented reality and telemedicine. A possible
stopgap measure would be to restrict access to this timer to applications
which gain the user's consent (for example, by displaying a confirmation
window) or the approval of some third party (for example, by being
downloaded from a trusted ``app store''). 

An interesting approach could be the use of heuristic profiling to
detect and prevent this specific kind of attack. Just like the abundance
of arithmetic and bitwise instructions was used by Wang et al. to
indicate the existence of cryptographic primitives \cite{DBLP:conf/esorics/WangJCWG09},
it can be noted that the various measurement steps of our attack access
memory in a very particular pattern. Since modern Javascript runtimes
already scrutinize the runtime performance of code as part of their
profile-guided optimization mechanisms, it should be possible for
the Javascript runtime to detect profiling-like behavior from executing
code and then modify its response accordingly (for example by jittering
the high-resolution timer, dynamically moving arrays around in memory,
etc).

\subsection{Conclusion}

In this report, we showed how the micro-architectural side-channel
attack, which is already recognized as an extremely potent attack
method, can be effectively launched from an untrusted web page. Instead
of the traditional cryptanalytic application of the cache attack,
we instead showed how user behavior can be effectively tracked using
this method. The potential reach of side-channel attacks has been
extended, meaning that additional classes of secure systems must be
designed with side-channel countermeasures in mind.

\subsection*{Acknowledgements}

We are thankful to Henry Wong for his investigation of the Ivy Bridge
cache replacement policy and to Burton Rosenberg for his tutorial
about pages and page frames.

\bibliographystyle{plain}
\bibliography{peeping}

\begin{thebibliography}{10}

\bibitem{StatCounter}
Statcounter globalstats.
\newblock Online, January 2015.
\newblock \url{http://gs.statcounter.com}.

\bibitem{DBLP:conf/ccs/Aciicmez07}
Onur Acii{\c{c}}mez.
\newblock Yet another microarchitectural attack: : exploiting i-cache.
\newblock In Peng Ning and Vijay Atluri, editors, {\em Proceedings of the 2007
  {ACM} workshop on Computer Security Architecture, {CSAW} 2007, Fairfax, VA,
  USA, November 2, 2007}, pages 11--18. {ACM}, 2007.

\bibitem{DBLP:conf/raid/ApececheaIES14}
Gorka~Irazoqui Apecechea, Mehmet~Sinan Inci, Thomas Eisenbarth, and Berk Sunar.
\newblock Wait a minute! {A} fast, cross-vm attack on {AES}.
\newblock In Angelos Stavrou, Herbert Bos, and Georgios Portokalidis, editors,
  {\em Research in Attacks, Intrusions and Defenses - 17th International
  Symposium, {RAID} 2014, Gothenburg, Sweden, September 17-19, 2014.
  Proceedings}, volume 8688 of {\em Lecture Notes in Computer Science}, pages
  299--319. Springer, 2014.

\bibitem{djb-cache}
Daniel~J. Bernstein.
\newblock Cache-timing attacks on {AES}.
\newblock Online, November 2004.
\newblock \url{http://cr.yp.to/papers.html#cachetiming}.

\bibitem{DBLP:conf/sp/ChenWWZ10}
Shuo Chen, Rui Wang, XiaoFeng Wang, and Kehuan Zhang.
\newblock Side-channel leaks in web applications: {A} reality today, a
  challenge tomorrow.
\newblock In {\em 31st {IEEE} Symposium on Security and Privacy, S{\&}P 2010,
  16-19 May 2010, Berleley/Oakland, California, {USA}}, pages 191--206. {IEEE}
  Computer Society, 2010.

\bibitem{javascript-api}
World Wide~Web Consortium.
\newblock Javascript {APIs}.
\newblock Online.
\newblock \url{http://www.w3.org/standards/techs/js}.

\bibitem{ecma-262}
{ECMA}.
\newblock Standard {ECMA-262}: {ECMAScript} language specification.
\newblock Online, June 2011.
\newblock
  \url{http://www.ecma-international.org/publications/standards/Ecma-262.htm}.

\bibitem{DBLP:conf/crypto/EisenbarthKMPSS08}
Thomas Eisenbarth, Timo Kasper, Amir Moradi, Christof Paar, Mahmoud
  Salmasizadeh, and Mohammad T.~Manzuri Shalmani.
\newblock On the power of power analysis in the real world: {A} complete break
  of the keeloqcode hopping scheme.
\newblock In David Wagner, editor, {\em Advances in Cryptology - {CRYPTO} 2008,
  28th Annual International Cryptology Conference, Santa Barbara, CA, USA,
  August 17-21, 2008. Proceedings}, volume 5157 of {\em Lecture Notes in
  Computer Science}, pages 203--220. Springer, 2008.

\bibitem{TypedArray}
Khronos Group.
\newblock Typed array specification.
\newblock Online, July 2013.
\newblock \url{https://www.khronos.org/registry/typedarray/specs/latest/}.

\bibitem{DBLP:conf/woot/HoBBP14}
Grant Ho, Dan Boneh, Lucas Ballard, and Niels Provos.
\newblock Tick tock: Building browser red pills from timing side channels.
\newblock In Sergey Bratus and Felix F.~X. Lindner, editors, {\em 8th {USENIX}
  Workshop on Offensive Technologies, {WOOT} '14, San Diego, CA, USA, August
  19, 2014.} {USENIX} Association, 2014.

\bibitem{DBLP:conf/sp/Hu92}
Wei{-}Ming Hu.
\newblock Lattice scheduling and covert channels.
\newblock In {\em 1992 {IEEE} Computer Society Symposium on Research in
  Security and Privacy, Oakland, CA, USA, May 4-6, 1992}, pages 52--61. {IEEE}
  Computer Society, 1992.

\bibitem{DBLP:conf/sp/HundWH13}
Ralf Hund, Carsten Willems, and Thorsten Holz.
\newblock Practical timing side channel attacks against kernel space {ASLR}.
\newblock In {\em 2013 {IEEE} Symposium on Security and Privacy, {SP} 2013,
  Berkeley, CA, USA, May 19-22, 2013}, pages 191--205. {IEEE} Computer Society,
  2013.

\bibitem{DBLP:conf/sp/JanaS12a}
Suman Jana and Vitaly Shmatikov.
\newblock Memento: Learning secrets from process footprints.
\newblock In {\em {IEEE} Symposium on Security and Privacy, {SP} 2012, 21-23
  May 2012, San Francisco, California, {USA}}, pages 143--157. {IEEE} Computer
  Society, 2012.

\bibitem{DBLP:conf/crypto/Kocher96}
Paul~C. Kocher.
\newblock Timing attacks on implementations of diffie-hellman, rsa, dss, and
  other systems.
\newblock In Neal Koblitz, editor, {\em Advances in Cryptology - {CRYPTO} '96,
  16th Annual International Cryptology Conference, Santa Barbara, California,
  USA, August 18-22, 1996, Proceedings}, volume 1109 of {\em Lecture Notes in
  Computer Science}, pages 104--113. Springer, 1996.

\bibitem{DBLP:books/daglib/0017272}
Stefan Mangard, Elisabeth Oswald, and Thomas Popp.
\newblock {\em Power analysis attacks - revealing the secrets of smart cards}.
\newblock Springer, 2007.

\bibitem{w3c-time}
Jatinder Mann.
\newblock High resolution time.
\newblock W3C Recommendation, December 2012.
\newblock \url{http://www.w3.org/TR/hr-time/}.

\bibitem{DBLP:conf/ctrsa/OsvikST06}
Dag~Arne Osvik, Adi Shamir, and Eran Tromer.
\newblock Cache attacks and countermeasures: The case of {AES}.
\newblock In David Pointcheval, editor, {\em Topics in Cryptology - {CT-RSA}
  2006, The Cryptographers' Track at the {RSA} Conference 2006, San Jose, CA,
  USA, February 13-17, 2006, Proceedings}, volume 3860 of {\em Lecture Notes in
  Computer Science}, pages 1--20. Springer, 2006.

\bibitem{DBLP:conf/ches/OswaldP11}
David Oswald and Christof Paar.
\newblock Breaking mifare desfire {MF3ICD40:} power analysis and templates in
  the real world.
\newblock In Bart Preneel and Tsuyoshi Takagi, editors, {\em Cryptographic
  Hardware and Embedded Systems - {CHES} 2011 - 13th International Workshop,
  Nara, Japan, September 28 - October 1, 2011. Proceedings}, volume 6917 of
  {\em Lecture Notes in Computer Science}, pages 207--222. Springer, 2011.

\bibitem{2005-percival-cache}
Colin Percival.
\newblock {Cache missing for fun and profit}.
\newblock Online, 2005.
\newblock \url{http://www.daemonology.net/hyperthreading-considered-harmful/}.

\bibitem{DBLP:conf/ccs/RistenpartTSS09}
Thomas Ristenpart, Eran Tromer, Hovav Shacham, and Stefan Savage.
\newblock Hey, you, get off of my cloud: exploring information leakage in
  third-party compute clouds.
\newblock In Ehab Al{-}Shaer, Somesh Jha, and Angelos~D. Keromytis, editors,
  {\em Proceedings of the 2009 {ACM} Conference on Computer and Communications
  Security, {CCS} 2009, Chicago, Illinois, USA, November 9-13, 2009}, pages
  199--212. {ACM}, 2009.

\bibitem{DBLP:conf/esorics/WangJCWG09}
Zhi Wang, Xuxian Jiang, Weidong Cui, Xinyuan Wang, and Mike Grace.
\newblock Reformat: Automatic reverse engineering of encrypted messages.
\newblock In Michael Backes and Peng Ning, editors, {\em Computer Security -
  {ESORICS} 2009, 14th European Symposium on Research in Computer Security,
  Saint-Malo, France, September 21-23, 2009. Proceedings}, volume 5789 of {\em
  Lecture Notes in Computer Science}, pages 200--215. Springer, 2009.

\bibitem{DBLP:conf/uss/YaromF14}
Yuval Yarom and Katrina Falkner.
\newblock {FLUSH+RELOAD:} {A} high resolution, low noise, {L3} cache
  side-channel attack.
\newblock In Kevin Fu and Jaeyeon Jung, editors, {\em Proceedings of the 23rd
  {USENIX} Security Symposium, San Diego, CA, USA, August 20-22, 2014.}, pages
  719--732. {USENIX} Association, 2014.

\bibitem{Yarom_LGHL_15}
Yuval Yarom, Fangfei Liu, Qian Ge, Gernot Heiser, and Ruby~B. Lee.
\newblock Last-level cache side-channel attacks are practical.
\newblock In {\em IEEE Symposium on Security and Privacy (S\&P)}, San Jose, CA,
  US, may 2015.

\bibitem{DBLP:conf/osdi/ZeldovichBKM06}
Nickolai Zeldovich, Silas Boyd{-}Wickizer, Eddie Kohler, and David
  Mazi{\`{e}}res.
\newblock Making information flow explicit in histar.
\newblock In Brian~N. Bershad and Jeffrey~C. Mogul, editors, {\em 7th Symposium
  on Operating Systems Design and Implementation {(OSDI} '06), November 6-8,
  Seattle, WA, {USA}}, pages 263--278. {USENIX} Association, 2006.

\bibitem{DBLP:conf/ccs/ZhangJRR12}
Yinqian Zhang, Ari Juels, Michael~K. Reiter, and Thomas Ristenpart.
\newblock Cross-vm side channels and their use to extract private keys.
\newblock In Ting Yu, George Danezis, and Virgil~D. Gligor, editors, {\em the
  {ACM} Conference on Computer and Communications Security, CCS'12, Raleigh,
  NC, USA, October 16-18, 2012}, pages 305--316. {ACM}, 2012.

\bibitem{DBLP:conf/ccs/ZhangJRR14}
Yinqian Zhang, Ari Juels, Michael~K. Reiter, and Thomas Ristenpart.
\newblock Cross-tenant side-channel attacks in paas clouds.
\newblock In Gail{-}Joon Ahn, Moti Yung, and Ninghui Li, editors, {\em
  Proceedings of the 2014 {ACM} {SIGSAC} Conference on Computer and
  Communications Security, Scottsdale, AZ, USA, November 3-7, 2014}, pages
  990--1003. {ACM}, 2014.

\end{thebibliography}

\end{document}